\documentclass[prx,reprint,superscriptaddress,twocolumn,showkeys]{revtex4-1}
\usepackage[utf8]{inputenc}
\usepackage{amsmath}
\usepackage{braket}
\usepackage{graphicx}
\usepackage{pgfplots}
\pgfplotsset{compat=1.15}
\usepackage{csquotes}
\usepackage{hhline}
\usepackage[ruled,vlined]{algorithm2e}
\usepackage{amssymb}

\usepackage{dcolumn}
\usepackage{tabularx}
\usepackage[colorlinks=true,linkcolor=blue,citecolor=blue,urlcolor=blue]{hyperref}
\usepackage{braket}
\usepackage{float}
\usepackage{listings}
\usepackage{color}
\usepackage{subcaption}
\usepackage{placeins}
\definecolor{mygreen}{rgb}{0,0.6,0}
\definecolor{mygray}{rgb}{0.5,0.5,0.5}
\definecolor{mymauve}{rgb}{0.58,0,0.82}

\newcolumntype{C}{>{\centering\arraybackslash}X}
\begin{document}
%
\title{Quantum Simulation of Hawking Radiation Using VQE Algorithm on IBM Quantum Computer}
\author{Ritu Dhaulakhandi}
\email{ritudhaulakhandi3626@gmail.com}
\affiliation{Department of Physics, Indian Institute of Science Education and Research, Pune, 411008, Maharastra, India}

\author{Bikash K. Behera}
\email{bikas.riki@gmail.com}
\affiliation{Bikash's Quantum (OPC) Pvt. Ltd., Balindi, Mohanpur 741246, West Bengal, India}

\begin{abstract}
\textbf{Abstract} - Quantum computers have an exponential speed-up advantage over classical computers. One of the most prominent utilities of quantum computers is their ability to study complex quantum systems in various fields using quantum computational algorithms. Quantum computational algorithms \cite{qgc_KitaevRMS1997} can be used to study cosmological systems and how they behave with variations in the different parameters of the system. Here, we use the variational quantum eigensolver (VQE) \cite{qgc_PeruzzoPRD2014} algorithm to simulate the Hawking radiation \cite{qgc_HawkingN1974} phenomenon. VQE algorithm is a combination of quantum and classical computation methods used to obtain the minimum energy eigenvalue for a given Hamiltonian. Three different custom ansatzes are used in the VQE algorithm from which the results for the case with minimum errors are studied. We obtain the plots for temperature and power from the minimum energy eigenvalue recorded for different values of mass and distance from the center of the black hole. The final result is then analyzed and compared against already existing data on Hawking radiation.
\end{abstract}


\maketitle
\section{Introduction \label{qgc_Sec1}}

The field of quantum computing was first brought to light by the works of Benioff and Manin in 1980, Feynman in 1982, and Deutsch in 1985. Benioff \cite{qgc_BenioffPRL1982} gave a quantum mechanical model of the Turing machines, around the same time, Manin \cite{qgc_ManinSR1980} wrote the ingenious paper `computable and uncomputable' which caught the attention of many quantum physicists. Feynman \cite{qgc_FeynmanIJTP1982} later proposed to use quantum computers to simulate models that weren't possible on classical computers due to various factors. Further contributions by Deutsch \cite{qgc_DeutschPRSL1985} in the field of quantum computing revolutionized the idea of quantum computation. The idea of quantum computing still wasn't popular until a special algorithm came up a decade later. The development of Shor's algorithm \cite{qgc_ShorSIAMJSSC1997} in 1994 finally ignited an interest in quantum computing and its potential. It soon became a race to find new algorithms and applications to solve problems faster.

The development of quantum computers gave rise to a new era in science and technology. A quantum computer is a computer that performs computation using the principles of quantum mechanics \cite{qgc_GriffithsPPH2004}, increasing the computational power which wasn't procurable by a classical computer. Quantum computers are widely used for modeling quantum systems and solving problems using various algorithms \cite{qgc_KitaevRMS1997}. It has put in motion a large number of applications in various fields of physics \cite{qgc_BeheraQIP12019,qgc_Raj20,qgc_PalRG2018}. As mentioned earlier, Feynman was the one who proposed the use of a quantum computer to solve problems in quantum mechanics. He mainly focused on the use of quantum computers to perform simulations of various models since a classical computer does not have enough space to perform the simulation \cite{qgc_IngallsWSC2008}.

Quantum simulations \cite{qgc_GeorgescuRMP2014} of various systems are now performed using quantum computers. Any general Hamiltonian of a system is hard to simulate in a classical computer as they grow exponentially in size. Quantum simulation enables us to study how a quantum system behaves as it evolves with respect to certain parameters. It has many applications in areas of physics such as quantum cosmology \cite{qgc_LaflammePRD1987,qgc_AmsterdamskiPRD1985,qgc_PedramPLB2009,qgc_HawkingNPB1984,qgc_CapozzielloIJMP1993,qgc_KocherIEEE2018,qgc_HartlePRD1983,qgc_GangulyarXiv2019}, and condensed matter physics \cite{qgc_KunpjQI2020}. In this paper, we work with the variational quantum eigensolver algorithm to simulate Hawking radiation on the IBMQ experience \cite{qgc_IBM} platform. IBMQ experience enables us to perform quantum computation tasks from remote areas around the globe.

Hawking radiation \cite{qgc_HawkingN1974} is theoretical radiation that arises due to quantum effects near the black hole event horizon. Hawking predicted in 1974 that particle-antiparticle pairs created near the event horizon, that come into existence due to the intense force of gravity, sometimes results in one particle escaping into the universe while the other particle gets absorbed by the black hole. The black hole loses energy to compensate for the separation of the pairs it had created. This energy loss by the black hole gets detected as Hawking radiation. Many attempts have been made to find experimental evidence of Hawking radiation \cite{qgc_RubinoNJP2011}. Steinhauer created an analogous black hole \cite{qgc_SteinhauerNP2016} in his lab with the help of laser and cold rubidium atoms. The model resulted in the detection of only one frequency, which puts doubt on how close it is to the actual Hawking radiation theory.

We apply the variational quantum eigensolver(VQE) algorithm \cite{qgc_PeruzzoPRD2014} from the Qiskit Aqua library \cite{qgc_QAL} to find the eigenvalues of the Hamiltonian of a non-rotating black hole. The library also provides an exact eigensolver, which helps us to compare the results obtained from the VQE algorithm. We use the Schwarzschild metric and principle of least action to formulate the Hamiltonian required for the algorithm. We further obtain the Hamiltonian eigenvalues using the VQE algorithm to make plots of temperature, and power and study their variation with mass and distance from the center of the black hole. Two qubits are used to simulate each dimension of the Hamiltonian.

The rest of the paper is divided into the following sections: In Section \ref{qgc_Sec2}, we briefly discuss the black hole metric and how to obtain the Hamiltonian for the non-rotating black hole. In Section \ref{qgc_Sec3}, we discuss how to work with the obtained continuous Hamiltonian in the discrete space system. In Section \ref{qgc_Sec4}, we explain the VQE algorithm used to find the ground state energy value. In Section \ref{qgc_Sec5}, we talk about Hawking radiation and discuss the results obtained from the VQE algorithm from three different ansatzes.

\section{Black Hole Metric \label{qgc_Sec2}}
From Einstein's theory of general relativity, the solution of the Einstein field equations \cite{qgc_StephaniCUP2003} for a gravitational field of mass having spherical symmetry is the Schwarzschild metric. Spherical symmetry implies that we have three killing vector fields that are linearly independent. These vector fields are tangent to the spherical mass. When the spherical symmetry is applied to Einstein's field equation, and from the generalized Birkhoff's theorem \cite{qgc_BirkhoffHUP1923}, which states that the only vacuum solution with spherical symmetry is the Schwarzschild solution \cite{qgc_ReissnerADP1916,qgc_NordstromPNAW1918,qgc_DrosteKAVWSTA1916}, we find that the Schwarzschild metric which is given as:

\begin{equation}
g=-c^2d\tau^2=-(1-\frac{r_s}{r})c^2dt^2+(1-\frac{r_s}{r})^{-1}dr^2+r^2(d\Phi^2)
\end{equation}
where $d\Phi^2$=$d\theta^2+\sin{\theta}^2d\phi^2$, and $r_s$=$\frac{2GM}{c^2}$($G$ is the gravitational constant and $M$ is the mass of the black hole).

We need to map the above equation to the rectangular coordinate system in order to implement the computational model that is discussed in the next section. We use the following coordinate transformation to obtain the Schwarzschild metric in the rectangular coordinate system:

\begin{equation}
\begin{split}
x=r\sin{\theta}\cos{\phi}\\
y=r\sin{\theta}\sin{\phi}\\
z=r\cos{\theta}
\end{split}
\end{equation}
where $r$=$\sqrt{x^2+y^2+z^2}$

On performing the above transformation with some additional assumptions such as zero electric charge, zero angular momentum, and zero value of cosmological constant, we get the Schwarzschild metric in the rectangular coordinate system for a non-rotating black hole (we will be working in Planck units throughout the paper), \cite{qgc_KohliarXiv2011} which is given as follows:

\begin{equation}
\label{qgc_Eq3}
ds^2=-\frac{(1-GM/2r)^2}{(1+GM/2r)^2}dt^2+(1+GM/2r)^4(dx^2+dy^2+dz^2)
\end{equation}

We can break the above equation into space and time form ($3+1$) where the spatial 3-metric $\gamma_{ij}$ is given as:

\begin{equation}
\label{qgc_Eq4}
\gamma_{ij}=(1+\frac{GM}{2r})^4(dx^2+dy^2+dz^2)
\end{equation}

We observe that the 3-metric $\gamma_{ij}$ is independent of time, which means that the metric tensor remains constant throughout the time evolution. Therefore, we can say that the Hamiltonian remains constant with time. With the help of Jacobi's formulation of the principle of least action \cite{qgc_GoldsteinC2001} and inverse of Eq. \eqref{qgc_Eq4}, we can define the Hamiltonian for a non-rotating black hole as follows:

\begin{equation}
\label{qgc_Eq5}
H^d=\frac{1}{2}(1+\frac{GM}{2r})^{1/4}[\frac{(p^d_x)^2}{2}+\frac{(p^d_y)^2}{2}+\frac{(p^d_z)^2}{2}]  
\end{equation}

Our task is now to convert the Hamiltonian given in Eq. \eqref{qgc_Eq5} into discrete space Hamiltonian in order to obtain it in matrix form and use it in the VQE algorithm.

\section{Discretization of space \label{qgc_Sec3}}

The Hamiltonian obtained for the non-rotating black holes (Eq. \eqref{qgc_Eq5}) has a continuous eigenspectrum. To simulate it on a quantum computer, we need to map it to a discrete space system \cite{qgc_JainRG2019}. We discretize it into two dimensional space with $x$, $y$ $\epsilon$ $[-L,L]$, such that $x$, $y$ each has $N$ eigenvalues. We finally obtain a mesh with $N^2$ spatial elements, where each element corresponds to an eigenvalue specific to the $x$ and $y$ value for that element. Let the mesh be centered at $[0,0]$ then we have an $N \times N$ matrix for the position operator \cite{qgc_SommaQIC2016}, with position eigenvalues lying along the diagonal. The position operator is given as:

\begin{eqnarray}
   x^d =\sqrt{\frac{\pi}{2N}} \begin{bmatrix}
   -N/2&0&0& &.&0 \\
   0&(-N/2)+1&0& &.&0 \\
   0&.&.& &.&. \\
   0&.&.& &.&.\\
   .&.&.& &.&.\\
   0&0&0& &.&(N/2)-1\\
   \end{bmatrix}
   \label{qnm_Eqn6}
\end{eqnarray}

To obtain the momentum operator, we take the discrete Fourier transform \cite{qgc_QFTWiki} of the discrete position operator. The wave function is now in the momentum space where the momentum operator works multiplicatively and the momentum eigenvalues are the same as the position eigenvalues for corresponding discrete space points. We then perform an inverse Fourier transform to bring the wave function back to discrete position space. The momentum operator is given as follows:
\begin{equation}
p^d=(F^d)^{-1} x^d F^d
\end{equation}

where $F^d$ is the discrete quantum Fourier transform matrix whose matrix elements are given as follows:
\begin{equation}
[F^d]_{j,k}=\frac{exp(i2\pi jk/N)}{N^{1/2}}
\end{equation}

Now that we have the position operator and the momentum operator, we just need to choose a value of $N$ and get the Hamiltonian in matrix form. We will work for the system with $N$=$4$. For higher values of $N$, the number of spatial points in the mesh will increase, the number of qubits required to perform the simulation will also increase according to the definition of the discrete space operators. The simulation will be closer to the actual model for very high values of $N$ and hence provide more accurate results. A system with a higher number of qubits requires more parameters in the ansatz to find the ground state wave function.

The discrete position and momentum operator for $N$=$4$ are given as:
\begin{eqnarray}
   x^d =\sqrt{\frac{\pi}{8}} \begin{bmatrix}
   -2&0&0&0 \\
   0&-1&0&0 \\
   0&0&0&0 \\
   0&0&0&1\\
   \end{bmatrix}
   \label{qnm_Eqn9}
\end{eqnarray}

\begin{eqnarray}
   p^d =\frac{\sqrt{\pi}}{8\sqrt{2}} \begin{bmatrix}
   -2&-2-2i&-2&-2+2i \\
   -2+2i&-2&-2-2i&-2 \\
   -2&-2+2i&-2&-2-2i \\
   -2-2i&-2&-2+2i&-2\\
   \end{bmatrix}
   \label{qnm_Eqn10}
\end{eqnarray}

We can now write down the discrete Hamiltonian for non-rotating black holes and use it to perform the simulation using the VQE algorithm. All the calculations performed so far only account for a single dimension. In order to extend it to the higher dimensions, we perform a tensor product of the Hamiltonian with identity matrices to include different dimensions, which is given as follows:
\begin{equation}
\label{qgc_Eq11}
H^d=\frac{1}{2}(1+\frac{GM}{2r})^{1/4}[(p^d)^2 \otimes I \otimes I+I \otimes (p^d)^2 \otimes I+I \otimes I\otimes (p^d)^2]
\end{equation}

We map the obtained Hamiltonian to Pauli basis $(\sigma_x,\sigma_y,\sigma_z,I)$ for the ease of implementation of the quantum circuit. The final expression for the Hamiltonian in Pauli basis is given as follows:

\begin{widetext}
\label{qgc_Eq12}
\begin{eqnarray}
H^d=\frac{1}{2}(1+\frac{GM}{2r})^{1/4}[\frac{\pi}{8}( \sigma_x \otimes \sigma_x \otimes I \otimes I )
+\frac{\pi}{16}( \sigma_x \otimes I \otimes I \otimes I )
+\frac{\pi}{8}( I \otimes \sigma_x \otimes I \otimes I )
+\frac{3\pi}{16}( I \otimes I \otimes I \otimes I )\nonumber\\
+\frac{\pi}{8}( I \otimes \sigma_x \otimes \sigma_x \otimes I )
+\frac{\pi}{16}(I \otimes \sigma_x \otimes I \otimes I )
+\frac{\pi}{8}(I \otimes I \otimes \sigma_x \otimes I )
+\frac{3\pi}{16}(I \otimes I \otimes I \otimes I )\nonumber\\
+\frac{\pi}{8}(I \otimes I\otimes \sigma_x \otimes \sigma_x )
+\frac{\pi}{16}(I \otimes I\otimes \sigma_x \otimes I )
+\frac{\pi}{8}(I \otimes I\otimes I \otimes \sigma_x )
+\frac{3\pi}{16}(I \otimes I\otimes I \otimes I )]
\end{eqnarray}
\end{widetext}

\section{Variational Quantum Eigensolver Algorithm \label{qgc_Sec4}}

\begin{figure}[]
\centering
\includegraphics[width=1\linewidth,height=4cm]{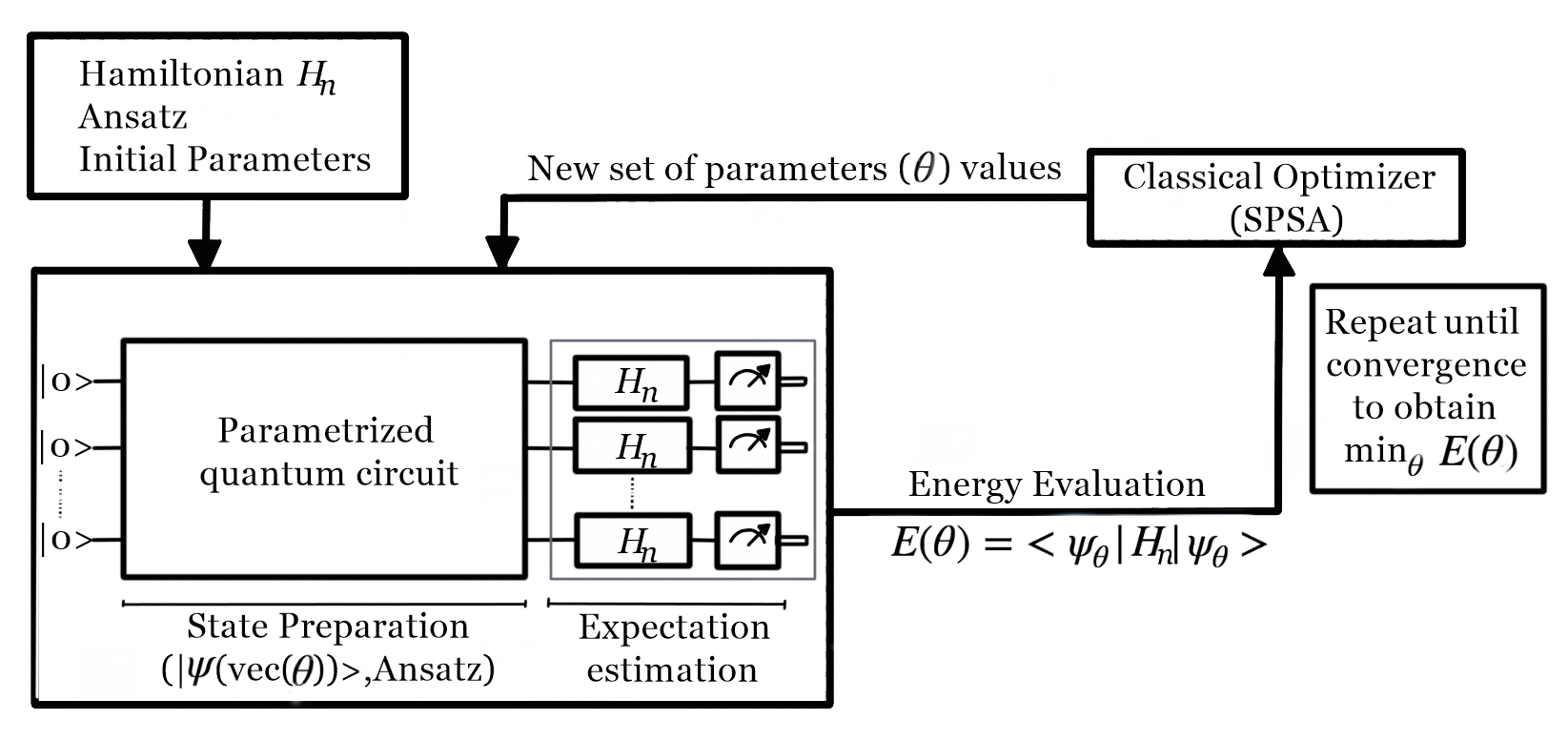}
\caption{This is a schematic diagram of how the VQE algorithm works. We feed in the Hamiltonian, ansatz that will be used to search for the state $\psi$, and initial parameters of the wave function. After measurement, we use a classical optimizer to update the parameters to find the minimum value of energy eigenvalue. The steps are iterated a large number of times to get $E_{min}$.}
\label{qfa_Fig1}
\end{figure}

A wave function carries information about the quantum state of the system and how it evolves with time. Furthermore, any normalized wave function can be written as a superposition of its eigenstates:
\begin{equation}
\label{qgc_Eq13}
\psi=\Sigma c_n\psi_n
\end{equation}

where $\big<\psi|\psi\big>=\sum|c_n|^2=1$ and $H\psi_n=E_n \psi_n$.
\\

We know that the ground state energy of a system $E_g$ is less than or equal to $E_n$. With the help of variational principle in quantum mechanics \cite{qgc_GriffithsPPH2004}, we can write the expectation value of the Hamiltonian as follows:

\begin{equation}
\label{Eqn14}
\big<H\big> =\big<\psi|H|\psi\big>=E_n\big<\psi|\psi\big>\geq E_g\sum|c_n|^2=E_g
\end{equation}
\\

In variational quantum eigensolver (VQE) \cite{qgc_PeruzzoPRD2014} algorithm, we first prepare an ansatz which is used to search for the ground state wave function ($\psi(\vec{\theta})$) for a given Hamiltonian. The initial parameters of the ansatz are taken to be random but they can be changed according to the user. Once we find the wave function, we apply the Hamiltonian operator and perform a measurement to obtain the energy eigenvalue ($E_n$). We then use a classical optimizer that updates the parameters ($\vec{\theta}$) to obtain an upper bound for the energy eigenvalue ($E_g$). The updated parameters are now used as the new parameters of the ansatz and the whole process is repeated unless the error value (the difference between the energy between two consecutive runs of the algorithm) in the classical optimizer is reached. Fig.~\ref{qfa_Fig1} gives a schematic description of the VQE algorithm. We use the simultaneous perturbation stochastic approximation optimizer (SPSA) \cite{qgc_SpallIEEE1992} to find the upper bound for energy eigenvalue for different values of $M$ and $r$ in the Hamiltonian.

\begin{algorithm}
\caption{VQE Algorithm}
\label{VQE algorithm}
\DontPrintSemicolon
\SetAlgoLined
\SetKwInOut{Input}{Input}\SetKwInOut{Output}{Output}
\BlankLine
\Input{Hamiltonian obtained after discretization\\
Ansatz to search for the ground state wave\\
function\\
Initial parameters to be used in the ansatz\\
Classical optimizer to be used\\
}
\Output{Upper bound for energy eigenvalue for given value of mass and distance from center of the black hole}
\BlankLine
The VQE algorithm follows the following steps after providing the input:\\
\BlankLine
\textbf{Step-1:} From the initial set of parameters of the\\ 
\hspace{1.2cm} ansatz we obtain an initial wave function.\\
\textbf{Step-2:} We measure the expectation value of the\\ 
\hspace{1.2cm} Hamiltonian operator using the initial wave\\
\hspace{1.2cm} function and obtain energy.\\

\textbf{Step 3:} The SPSA optimizer uses gradient descend\\
\hspace{1.2cm} method to obtain new set of parameters for\\
\hspace{1.2cm} the ansatz.\\

\textbf{Step 4:} Steps 1, 2, and 3 are repeated to get another\\ 
\hspace{1.2cm} value of energy and the parameters which\\
\hspace{1.2cm} gave closer energy value is used to generate\\
\hspace{1.2cm} new sets of parameters.\\

\textbf{Step 5:} The whole process is repeated until the error\\
\hspace{1.2cm} difference between two energy value obtained\\
\hspace{1.2cm} in two consecutive run is equal to error set\\ 
\hspace{1.2cm} in the optimizer.\\
\BlankLine
\end{algorithm}

\begin{figure*}[!ht]
\centering
\begin{subfigure}{0.3\linewidth}
\includegraphics[width=\linewidth]{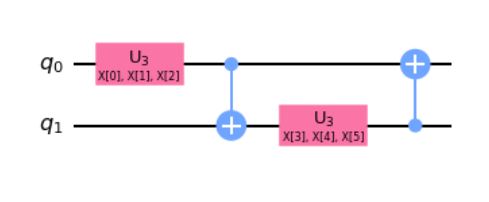} 
\caption{Ansatz 1}
\end{subfigure}\hfill
\begin{subfigure}{0.3\linewidth}
\includegraphics[width=\linewidth]{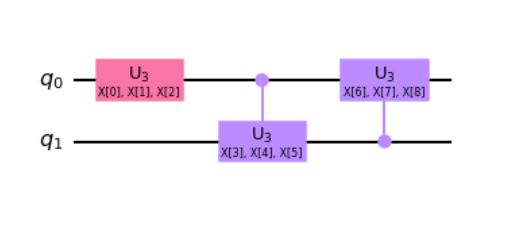} 
\caption{Ansatz 2}
\end{subfigure}\hfill
\begin{subfigure}{0.3\linewidth}
\includegraphics[width=\linewidth]{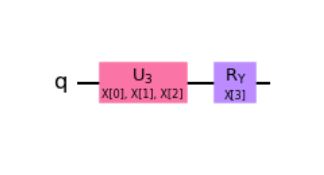} 
\caption{Ansatz 3}
\end{subfigure}\hfill
\caption{(a)Ansatz 1 consists of a combination of $U_3$ gates and $CNOT$ gates between two qubits as shown in the figure. This combination is repeated for all possible pairs of two qubits for systems with more number of qubits. For the VQE algorithm, only one iteration is used for our system. (b)Ansatz 2 is made up of $U_3$ gates and control $U_3$ gates between two qubits as shown. Similar to part (a) the combination is repeated for systems with more number qubits. Only one iteration is used for this circuit as well in the VQE algorithm. (c) Ansatz 3 consists of a $U_3$ gate and an $Ry$ gate. For our system, it is iterated twice in the VQE algorithm.}
\label{qdctc_Fig2}
\end{figure*}

\section{Hawking Radiation \label{qgc_Sec5}}

Hawking's revelation about black hole evaporation was based on a quantum physics phenomenon of virtual particles, and their behavior near the event horizon \cite{qgc_ParentaniS2011,qgc_HawkingCMP1975}. Near the black hole event horizon, these virtual pairs consist of photons. One of the photons is absorbed into the black hole while the other escapes into the universe beyond the event horizon. A wave carrying negative energy is created due to quantum effects when the pairs get separated. This wave enters the black hole which in turn reduces its total mass or energy, which results in the black hole evaporating over time.

From conservation laws and thermal properties of the black hole \cite{qgc_HawkingN1974,qgc_LoprestoPT2003}, Hawking was able to conclude that the black hole emits electromagnetic radiation whose temperature is inversely proportional to the mass of the black hole ($T=\frac{1}{M}$). The Bekenstein-Hawking luminosity \cite{qgc_LoprestoPT2003,qgc_PagePRD1976} further assumes that the emission only consists of photons and the horizon is taken as the radiating surface. From this, it was found that the luminosity ($P$) of a black hole is proportional to $\frac{1}{M^2}$.

\begin{figure*}[!ht]
\centering
\begin{subfigure}{0.3\linewidth}
\includegraphics[width=\linewidth]{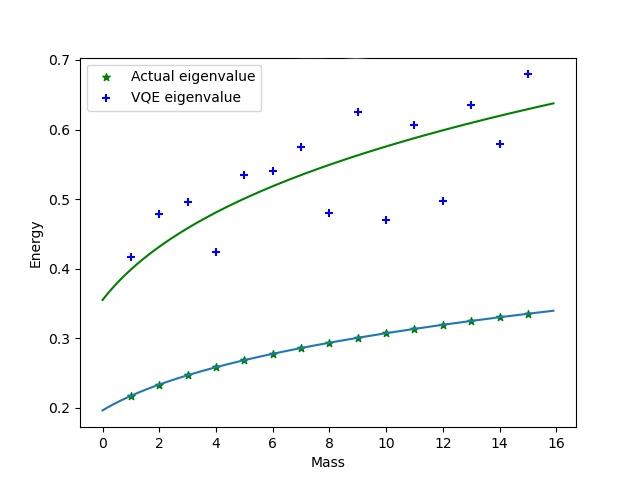} 
\caption{}
\end{subfigure}\hfill
\begin{subfigure}{0.3\linewidth}
\includegraphics[width=\linewidth]{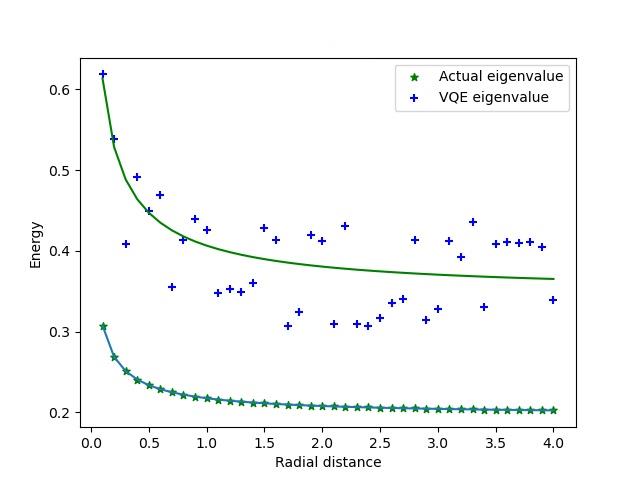} 
\caption{}
\end{subfigure}\hfill
\begin{subfigure}{0.3\linewidth}
\includegraphics[width=\linewidth]{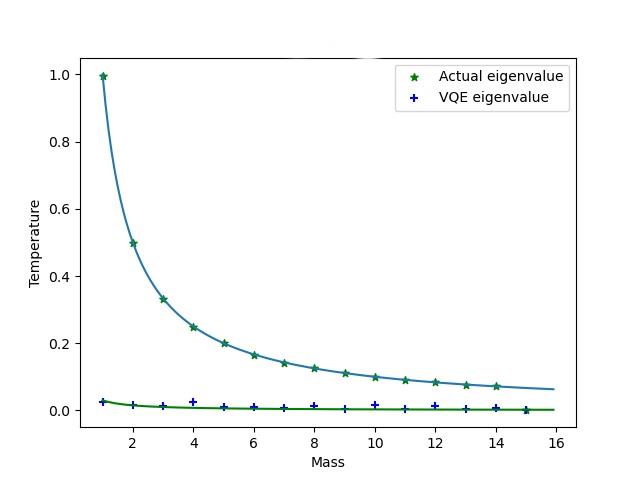} 
\caption{}
\end{subfigure}\hfill
\begin{subfigure}{0.3\linewidth}
\includegraphics[width=\linewidth]{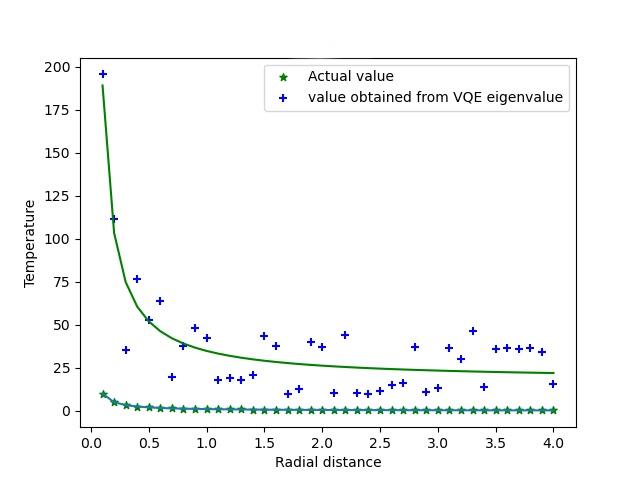} 
\caption{}
\end{subfigure}\hfill
\begin{subfigure}{0.3\linewidth}
\includegraphics[width=\linewidth]{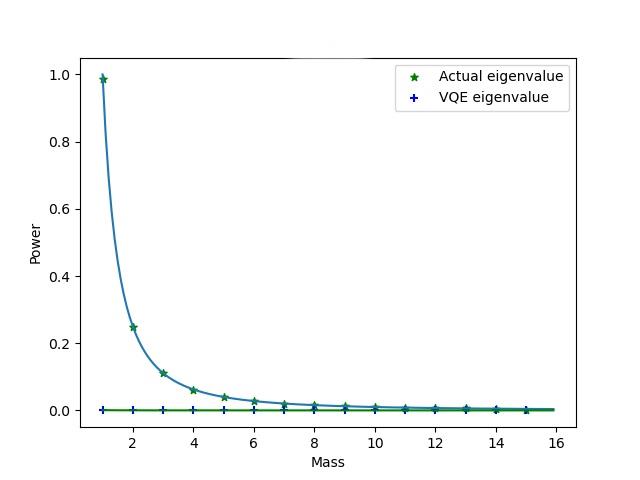} 
\caption{}
\end{subfigure}\hfill
\begin{subfigure}{0.3\linewidth}
\includegraphics[width=\linewidth]{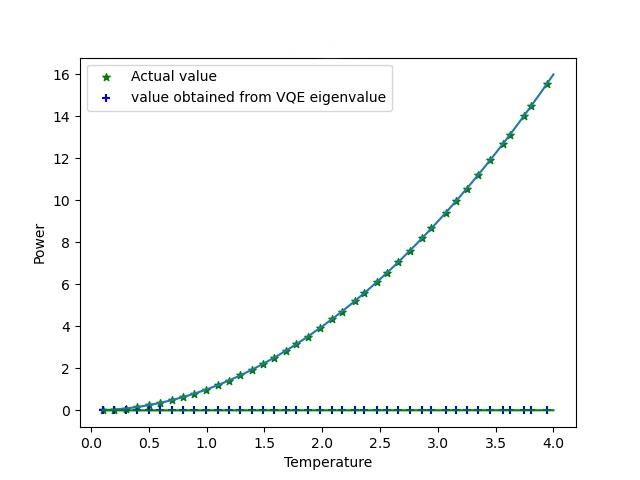} 
\caption{}
\end{subfigure}\hfill
\caption{\textbf{Ansatz 1 Results:} (a) The mass vs energy plot has an energy difference between the actual and the VQE data. There is also a large deviation in the data obtained from VQE. (b) In the radial distance vs energy plot, we can see a large amount of deviation in the VQE data along with the energy gap. (c) The deviation in the values of temperature has decreased but the gap between the actual and VQE data is huge at first and decreases to reach a closer value as mass value is increased. (d) The deviation in the temperature values is high and the gap is also huge at first and decreases as the value of the radial distance is increased. (e) There is a negligible deviation in the power values but there is a gap between actual and VQE values which decreases as the value of mass is increased. (f) Again the deviation in data values of power is negligible but the gap between the actual and VQE data increases with an increase in temperature.}
\label{qdctc_Fig3}
\end{figure*}

\begin{figure*}[!ht]
\centering
\begin{subfigure}{0.3\linewidth}
\includegraphics[width=\linewidth]{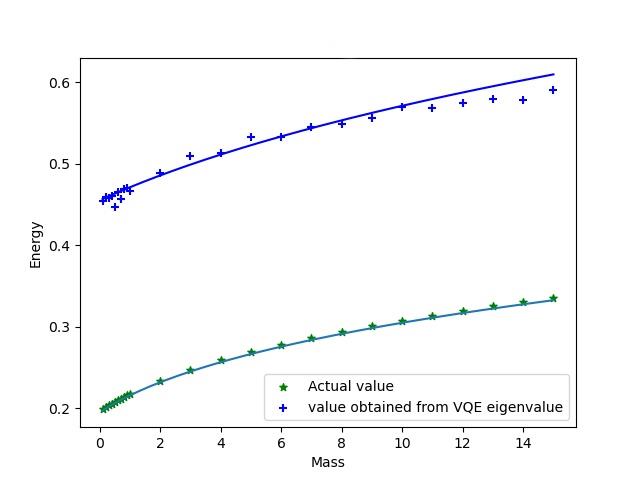} 
\caption{}
\end{subfigure}\hfill
\begin{subfigure}{0.3\linewidth}
\includegraphics[width=\linewidth]{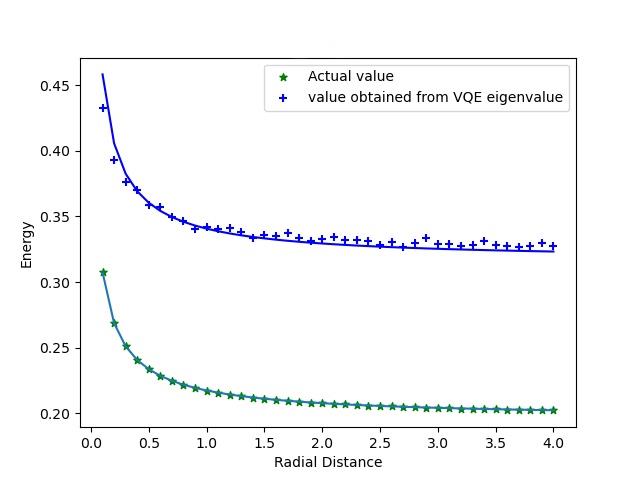} 
\caption{}
\end{subfigure}\hfill
\begin{subfigure}{0.3\linewidth}
\includegraphics[width=\linewidth]{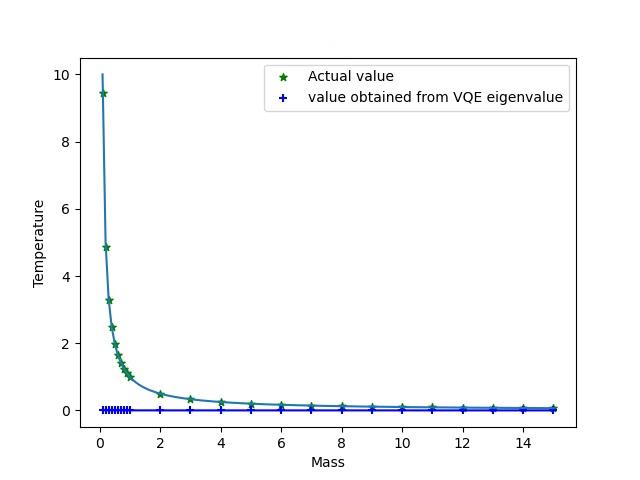} 
\caption{}
\end{subfigure}\hfill
\begin{subfigure}{0.3\linewidth}
\includegraphics[width=\linewidth]{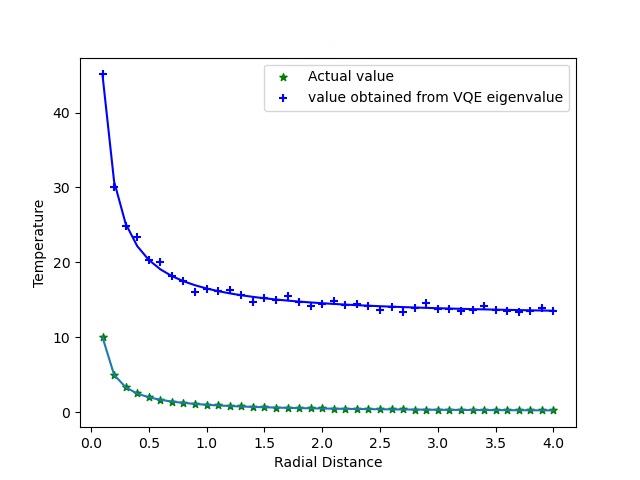} 
\caption{}
\end{subfigure}\hfill
\begin{subfigure}{0.3\linewidth}
\includegraphics[width=\linewidth]{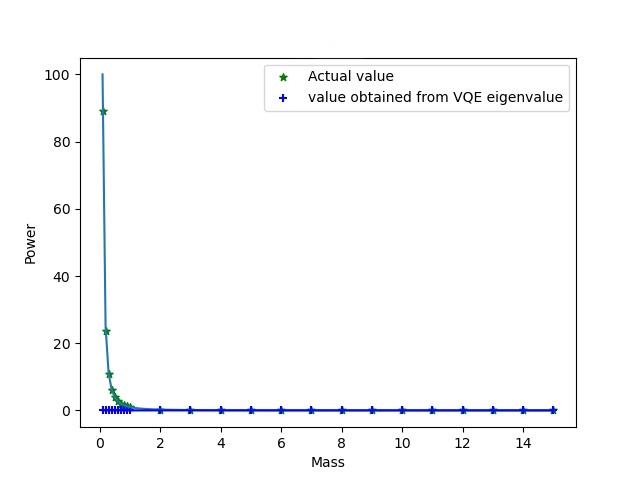} 
\caption{}
\end{subfigure}\hfill
\begin{subfigure}{0.3\linewidth}
\includegraphics[width=\linewidth]{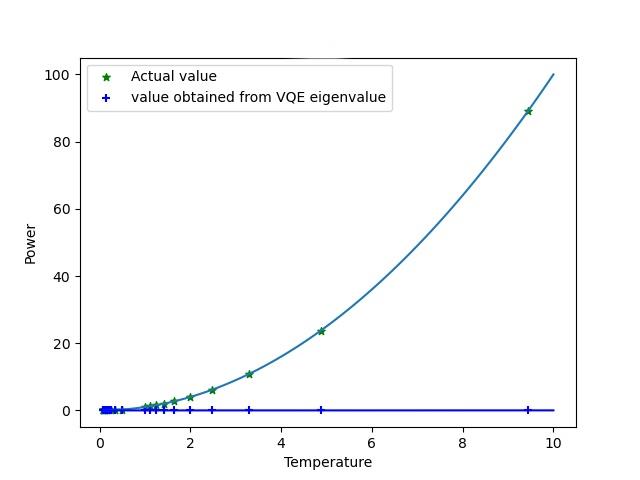} 
\caption{}
\end{subfigure}\hfill
\caption{\textbf{Ansatz 2 Results:} (a) For the given Ansatz there is an energy gap in the mass vs energy plot but the deviation in the VQE value is very small compared to Ansatz 1. (b) Again there is an energy gap between the actual and VQE data in the radial distance vs energy plot but deviation in VQE data is negligible. (c) The gap between the actual and VQE data obtained for temperature is very high for smaller values of mass and it decreases as mass value increases. (d) The gap between the two data of temperature is huge at the beginning but it decreases to a constant difference as the value of radial distance increases. (e) There is a huge difference in power values for the two data for a mass value less than one solar mass. For higher mass values the two data give almost the same values. (f) The gap between the actual and VQE data obtained for power increases with an increase in the value of temperature.}
\label{qdctc_Fig4}
\end{figure*}

\begin{figure*}[!ht]
\centering
\begin{subfigure}{0.3\linewidth}
\includegraphics[width=\linewidth]{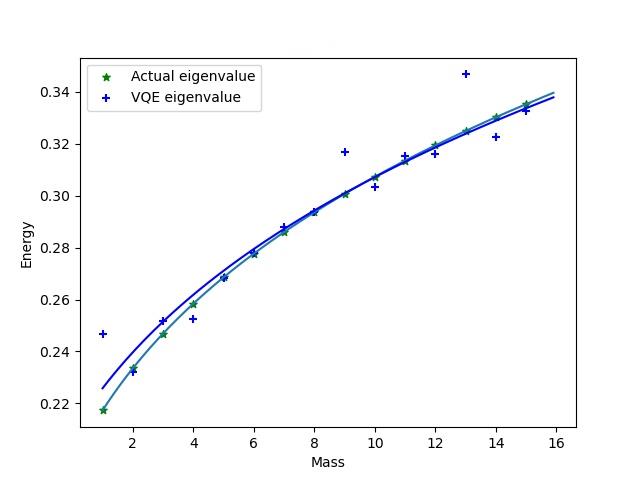} 
\caption{}
\end{subfigure}\hfill
\begin{subfigure}{0.3\linewidth}
\includegraphics[width=\linewidth]{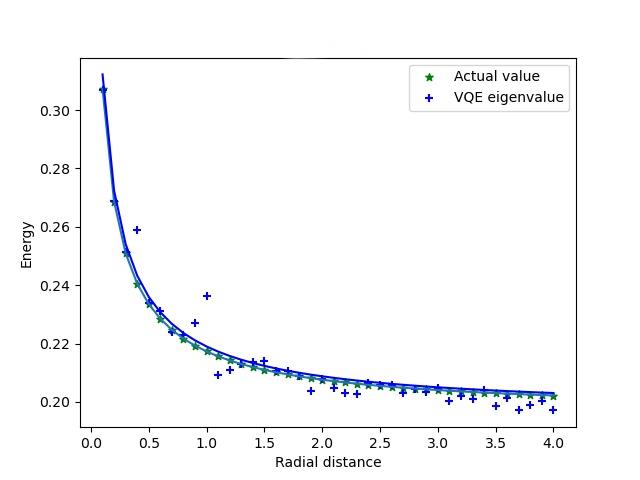} 
\caption{}
\end{subfigure}\hfill
\begin{subfigure}{0.3\linewidth}
\includegraphics[width=\linewidth]{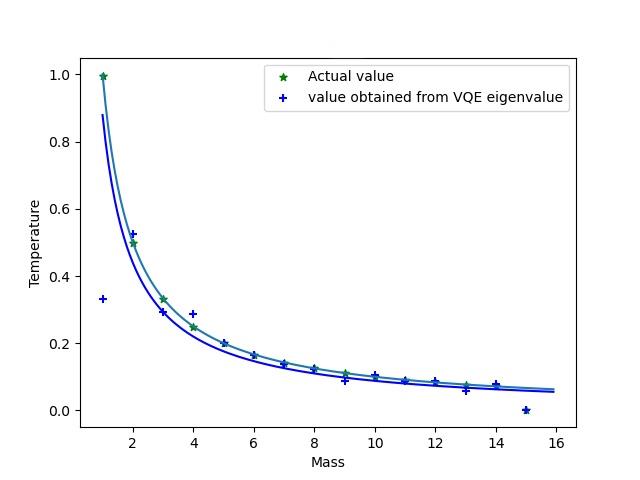} 
\caption{}
\end{subfigure}\hfill
\begin{subfigure}{0.3\linewidth}
\includegraphics[width=\linewidth]{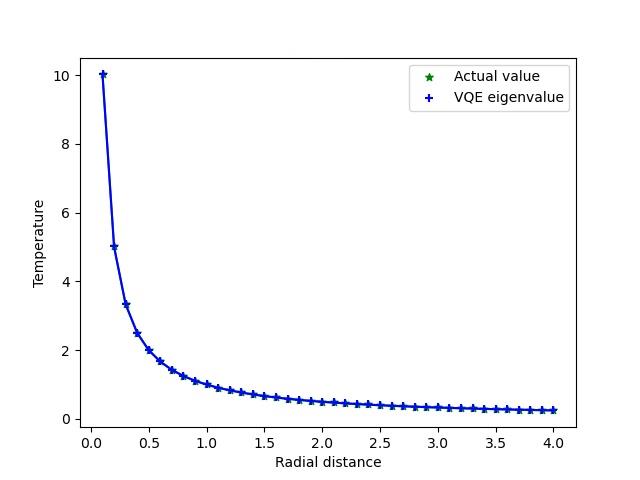} 
\caption{}
\end{subfigure}\hfill
\begin{subfigure}{0.3\linewidth}
\includegraphics[width=\linewidth]{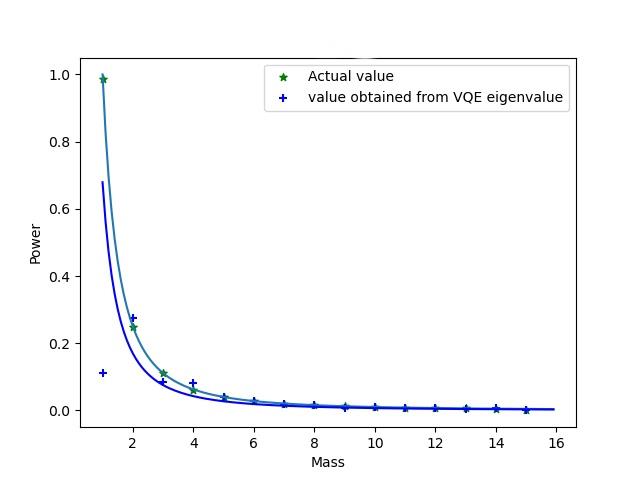} 
\caption{}
\end{subfigure}\hfill
\begin{subfigure}{0.3\linewidth}
\includegraphics[width=\linewidth]{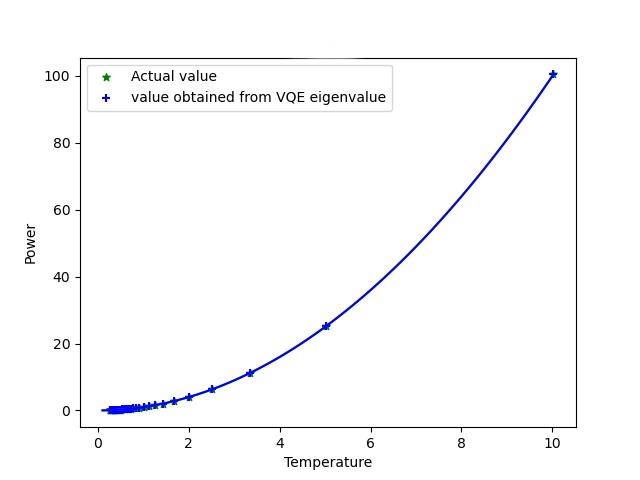} 
\caption{}
\end{subfigure}\hfill
\caption{\textbf{Ansatz 3 Results:} (a) The mass vs energy plot for the actual and VQE data are almost the same with few deviations in the VQE data. (b) Small deviations can be observed but overall the plots for actual and VQE data are close. (c) The mass vs temperature plot for actual and VQE data is almost the same with a small deviation in VQE data. (d) The plots of actual and VQE data are the same for radial distance vs temperature. (e) There are small deviations in VQE data for power from the actual data for different values of mass but the two plots are still close. (f) There is almost no error in the temperature vs power plot. The actual and VQE data plots have the same expression for power.}
\label{qdctc_Fig5}
\end{figure*}

We work with three different ansatzes in the VQE algorithm for the Hamiltonian for a non-rotating and non-charged Schwarzschild black hole obtained from Eq. \eqref{qgc_Eq13}. The $U_3$ and $CNOT$ gates used in ansatzes are given as follows:
\begin{eqnarray}
   U_3(\theta,\phi,\lambda) =\begin{bmatrix}
   \cos{\frac{\theta}{2}}&-e^{i\lambda}\sin{\frac{\theta}{2}} \\
   e^{i\phi}\sin{\frac{\theta}{2}}&e^{i\phi+i\lambda}\cos{\frac{\theta}{2}} \\
   \end{bmatrix}
   \label{qnm_Eqn15}
\end{eqnarray}
\begin{eqnarray}
   CNOT=\begin{bmatrix}
   1&0&0&0 \\
   0&1&0&0 \\
   0&0&0&1 \\
   0&0&1&0 \\
   \end{bmatrix}
   \label{qnm_Eqn16}
\end{eqnarray}

We then compare the results obtained for the three ansatzes and find the ansatz with the least error. The circuit used for the three ansatzes can be seen in Fig.~\ref{qdctc_Fig2}. Ansatz 1 has an energy difference of at least nearly 0.2 units from the exact value. It also had a higher variance from the fitted plot in VQE data than the other ansatzes. Ansatz 3 also has an energy difference from the exact value but the variation from the fitted plot in the VQE data is less than Ansatz 1. Ansatz 2 gives nearly the same result as the exact value. The energy gap could be there due to the use of control gates as we have not used any control gates in Ansatz 2. The variation in Ansatz 1 is probably because of the high number of parameters used in the circuit. Ansatzes 2 and 3 have a lesser number of parameters.

Some of the common aspects of Fig.~\ref{qdctc_Fig3}, \ref{qdctc_Fig4}, and \ref{qdctc_Fig5}:- for mass vs energy plot, mass is taken as constant times the solar mass, where the constant belongs to natural numbers. The energy of the black hole varies as $a(b+cM)^{1/4}$, where $M$ is the mass of the black hole and $a$, $b$, and $c$ are constants. In the radial distance vs energy plot, the relation between energy and the distance from the center of the black hole is found to be $E$=$a(b+c/r)^{1/4}$, where $r$ is the distance from the center of the black hole and $a$, $b$, and $c$ are constants. $r$ is varied in orders of $GM$ where $M$ is the mass of the black hole. For the mass vs temperature plot, the value of temperature was obtained by using the expression of mass in terms of energy for a constant value of $r$. We know that the temperature is inversely proportional to the mass of the black hole. Using the expression of $r$ before and keeping the energy constant we find that the temperature also decreases with distance as seen in radial distance vs temperature plot. We find the expression for power by using the equation obtained for mass and energy. In the mass vs power plot, we observe that the power radiated by a black hole decreases with mass. Finally, we obtain the plot for power vs temperature with the help of the equation of curves obtained from other plots. The plot is similar to the data from other sources.

For Fig.~\ref{qdctc_Fig3}, Ansatz 1 results in an energy gap between the actual value and the VQE value. The possible reason could be the use of control gates which impose conditions on the qubit system hence increasing the energy of the system. There is also a lot of deviation observed in the VQE data which is mostly because of over-fitting due to a large number of parameters. In mass vs energy plot, the equation of the curve of VQE data $E$=$a'(b+c'M)^{1/4}$ differs from the actual data $E$=$a(b+cM)^{1/4}$ only in the values of $a$ and $c$, where $a'$ and $c'$ are not equal to $a$ and $c$ respectively. For the radial distance vs energy plot, we again notice the issue of the energy gap between the two data and deviation in the VQE data. The equation found for the curves is $E$=$a(b+c/r)^{1/4}$ where the values of $a$ and $c$ are different for the two data sets. The expression for mass is obtained from $E$=$a(b+cM)^{1/4}$ and further used to find the temperature for the two data set. Since temperature is inversely proportional to mass, we get that temperature is inversely proportional to the fourth power of ground state energy. This leads to a decrease in deviation of data values because all energy values are less than one but it results in a high energy gap for lower values of energy or in other words lower values of mass. The same procedure is followed to get the temperature in terms of energy where energy varies with radial distance. For this plot, a large amount of deviation is observed in the VQE data as the temperature is now directly proportional to the energy. We also observe that after a certain radial distance the temperature gap becomes constant which is similar to the radial distance vs energy plot. The expression for power is again obtained using the mass vs energy plot equation. Since power is inversely proportional to the square of the mass, deviation in data has decreased but again the gap is huge up to a certain value of mass. Both the data tend to zero for very high values of mass. Using the expressions of temperature and power already obtained before, we have plotted the temperature vs power plot. For Ansatz 1, the deviation in VQE data is small but the gap in power value of the two data sets increases with an increase in temperature value. 

For Fig.~\ref{qdctc_Fig4}, Ansatz 2 also results in an energy gap between the actual value and the VQE value for possibly the same reason as Ansatz 1. The deviation in the VQE data is much lower than the VQE data of Ansatz 1, mostly because we use fewer parameters. Since there is only an energy gap in plots for mass vs energy and radial distance vs energy, the equations of the curves have significant differences only in the value of $a$. There is not much difference in the value of $c$, as the deviation in VQE data is not that large. The reason for the gap between the two data sets in the remaining plots is the same as the one explained for Ansatz 1.

For Fig.~\ref{qdctc_Fig5}, Ansatz 3 gives very close results to actual value data. It does not include any control gates. It also has fewer parameters in comparison to Ansatz 1. All the plots show a very small amount of deviation but in general, there is a very small difference in the equation of the curves of the actual value and VQE value data sets.

In all the three ansatzes, we notice that the energy of the black hole decreases at a faster rate inside the event horizon while it decreases at a slower pace outside the event horizon. A black hole that has more mass has more energy than a black hole that has less mass. We also notice that the black holes with higher mass are at a lower temperature than the black holes with smaller mass which matches with the theory. The temperature is seen to decrease very quickly near the black hole and then at a slower pace for greater distance. It can be observed that a black hole with a smaller mass radiates more power than a black hole with a higher mass. Finally, the power vs temperature plot is similar to the data from various sources.

\section{Conclusion \label{qgc_Sec6}}
To conclude, we first obtain the Hamiltonian of a non-rotating and non-charged black hole using solutions of Einstein's field equations and proper approximations. We transform the obtained Hamiltonian to use it in the discrete space system. We use the final Hamiltonian in Pauli basis to perform the simulation for Hawking radiation using the VQE algorithm. We tried out different ansatzes to find the ground state wave function and compared the results obtained from them. We varied the values for mass and distance from the center of the black hole and recorded the energy eigenvalues for each case. The different plots of temperature and power prepared from the energy eigenvalues obtained from the VQE algorithm are closest with the theoretical values for Ansatz 3. Hence, we can say that quantum computation algorithms can be used to study cosmological phenomena which are hard to simulate in labs. As future applications, black holes with different structures or compositions, for example, rotating and charged black holes can also be simulated using the VQE algorithm, and accordingly, observations can be obtained.

\section*{Acknowledgments}
\label{qlock_acknowledgments}
The authors acknowledge the support of the IBM quantum experience. The views expressed are those of the authors and do not reflect the official policy or position of IBM or the IBMQ experience team.


\end{document}